\documentclass[
,preprint
,showpacs
]{revtex4}
\begin{document}
\title{Vibrations of closed-shell Lennard-Jones icosahedral and cuboctahedral clusters and their 
effect on the cluster ground state energy}
\author{Antonio \v{S}iber}
\email{asiber@ifs.hr}
\affiliation{Institute of Physics, P.O. Box 304, 10001 Zagreb, Croatia}
\begin{abstract}
Vibrational spectra of closed shell Lennard-Jones icosahedral and cuboctahedral clusters are 
calculated for shell 
numbers between 2 and 9. Evolution of the vibrational density of states with the cluster shell
number is examined and differences between icosahedral and cuboctahedral clusters described. 
This enabled a quantum calculation of quantum ground state energies of the clusters 
in the quasiharmonic approximation and 
a comparison of the differences between the two types of clusters. It is demonstrated 
that in the quantum 
treatment, the closed shell icosahedral clusters binding energies differ from those of 
cuboctahedral clusters more than is the case in classical treatment.
\end{abstract}
\maketitle

\section{Introduction}
\label{sec:sec0}

The interest of the condensed matter community in cluster physics has been recently revived due to 
exciting technological possibilities offered by new materials in which clusters play the role of 
basic "building blocks" (see e.g. Ref. \cite{nanoassembly}).
Theoretical investigations of model clusters, i.e. assemblies of particles 
interacting via binary (mostly Lennard-Jones) 
potential has a long 
history \cite{Burton,Burton2,Xie,McGinty1,Northby1,Waal0,Doye1,Doye3,Waal1,Leary,Doye2,Lim,Meyer}. Most 
studies performed so far have concentrated on the search of 
the most stable configuration of $N$ particles. Typically, total potential energy of 
a cluster is written as a sum over binary interactions of all pairs of particles within the cluster. 
Then, a search for a set of coordinates (or configuration of the cluster) which minimize 
the potential energy is performed \cite{ananas}. Depending on the number of particles within a cluster, 
very different cluster shapes can be obtained following this procedure \cite{Doye3}. The approach just sketched can 
require formidable numerical optimization procedures \cite{Leary}. However, it is completely 
classical since it produces an absolute minimum in the potential energy, rather than the energy 
of a ground quantum state of the system which must be smaller in absolute value. The difference 
between the two values is the (quantum) zero-point energy of the cluster.

The clusters investigated in this article are known as closed shell clusters. These appear for 
"magic values" of $N$, and the specific sequence to be investigated here is 
$N=13,55,147,309,561,923...$. The characteristic sequence of 
"magic" numbers has also been observed experimentally (see e.g. Ref. \cite{Miehle}). Closed shell 
clusters of interest to 
the present article appear in two configurations, 
icosahedral and cuboctahedral (see Fig. \ref{fig:fig1}). These clusters can be thought of as being 
assembled by adding closed "shells" of atoms to a single atom located at the origin, i.e. 
the center of the cluster. The thus "assembled" structure can be characterized with the 
maximum shell number $n$ (see Fig. \ref{fig:fig1}). The two types of clusters have quite 
different geometries, but their common feature is that they have the same number of 
atoms for the given maximum shell number $n$ \cite{Waal0}. This is a very convenient 
feature which enables a direct comparison of the various physical properties of 
such clusters \cite{Xie,Lim}. Cuboctahedral clusters can also 
be visualized as pieces of a crystal with face-centered-cubic (FCC) packing. A FCC crystal 
can be obtained from cuboctahedral clusters in the limit of infinite shell number. The 
same is not true for icosahedral clusters which are therefore "noncrystal" \cite{Waal0}. Closed shell 
icosahedral configuration 
has been shown to have lower total potential energy for clusters 
smaller than about 9000 particles \cite{Xie}. The actual number of particles (or the shell number) 
\cite{Xie} above which cuboctahedral clusters have lower potential energy has been a subject of 
debate (see e.g. Refs. \cite{Xie,Waal1,Kakar}). This is not important for the purpose of this article since I am 
going to consider relatively small clusters (up to nine shells or 2869 atoms) for which 
icosahedral configuration is a classically 
preferred one. A plausible reason for the fact that the closed shell icosahedral (CSIC) configuration 
is classically more stable from the closed shell cuboctahedral (CSFCC) configuration is that the 
arrangement of particles 
which are on the cluster surface is "tighter" in the CSIC than in the CSFCC ordering of the cluster 
(see Fig. \ref{fig:fig1}). 

The aim of this article is twofold. First, a detailed microscopic calculation of cluster vibrations 
shall be performed, depending on the closed-shell cluster size. It is of interest to see 
whether the different cluster geometries reflect themselves in the cluster vibrational 
properties. The differences in vibrational properties could be exploited to 
discriminate between the different cluster geometries. Some early studies \cite{Burton,
Burton2,McGinty1} have dealt with the vibrational frequency spectra of clusters, but 
these studies were limited to quite small clusters and there were no attempts 
to compare the vibrational spectra pertaining to clusters with the different symmetries. 
The results should also be relevant to the studies of vibrations in nanoparticles deposited 
on substrates \cite{Patton}.

The Second aim is to reexamine the 
cluster stability from the point of 
quantum mechanics, i.e. to investigate the differences between the CSIC and CSFCC 
clusters with respect to their quantum ground state energy and temperature 
dependent vibrational entropy. The quantum approach is quite easy in the 
case in which the cluster dynamics can be adequately represented by harmonic vibrations, but 
it should be kept in mind that such treatment is adequate only for sufficiently low 
temperatures. This is the region of temperatures that is of interest to this article.
There are several points that come to mind regarding 
the second aim. First, one could assume that the fact that the surfaces of CSFCC clusters are 
less densely packed implies that their characteristic vibrations will be "softer", i.e. 
of lower frequency. Thus, zero-point energy of CSFCC clusters could be expected to be smaller, 
and the shift of the ground state energy from absolute minimum of the potential energy lesser 
than in CSIC clusters. This would suggest that quantum ground state energy may even be larger 
(in absolute units) in CSFCC clusters, which would promote them to thermodynamically preferred 
configurations of $N$ atoms at zero temperature, $T=0$ K. Second, even if the zero-point energy 
is not sufficiently different in the two cases, in order to find the thermodynamically preferred configuration at 
finite temperature one should minimize free energy of the system. As the cluster free energy 
depends on the features of the cluster vibrational spectrum and temperature, it may happen that at 
some finite temperature, $T_c$, the free energy of the CSFCC configuration becomes smaller 
than the free energy of CSIC configuration, even if 
its ground state, binding energy was smaller than in the CSIC configuration. Third, as vibrational frequencies depend on the mass of particles in the cluster, $M$, 
quantum effects and thermodynamical considerations must be mass dependent. Classically, the 
cluster stability considerations depend only on the binary interaction potential. Thus, any two 
clusters composed of different isotopes of the same element will always have the same 
configuration. Obviously, this needs not be the case in a quantum treatment and quantum 
effects are going to be larger for clusters composed of lighter particles. This article 
aims at examining the three points mentioned above. The results of the article are 
not directly applicable to real clusters, although the model of the cluster adopted could 
be used to describe noble gas clusters and could serve as a point 
of departure for setting up more complex interaction models \cite{Lim,Meyer}. There are some reservations, 
however, since it is known that the form of the binary potential employed in calculation 
can influence the cluster optimal 
shape and its dynamical properties \cite{Doye1,Lim}.

The article is organized as follows. In Section \ref{sec:sec1}, I shall briefly describe 
the adopted theoretical approach (quantum quasiharmonic approximation). Section \ref{sec:sec2} 
deals with the model clusters in which 
the particles interact via Lennard-Jones binary potentials. Vibrational spectra of 
CSIC and CSFCC clusters are calculated for shell numbers between two (55 atoms) and nine (2869 atoms).
The effects of mass and temperature on the Helmholtz free energy of clusters are considered 
on the example of clusters of Ne, Ar and Xe atoms. Section \ref{sec:sec3} summarizes and concludes 
the article.

\section{Prerequisites for the calculation of the vibrations and vibrational free energy of clusters}

\label{sec:sec1}

The dynamical behavior of a cluster is described by a set of coordinates 
$\{{\bf r}_1, {\bf r}_2, ..., {\bf r}_N \}$ which are treated as time dependent variables. 
The particles within the cluster are assumed to interact via a binary potential, $v$, which 
depends only on their relative positions, i.e. 
$v({\bf r}_i, {\bf r}_j) = v(|{\bf r}_i - {\bf r}_j|)$. The total potential energy of the cluster, $V_p$ 
is assumed to be given by a summation of binary interactions over all the pairs of particles 
in the cluster,
\begin{equation}
V_p (\{{\bf r}_1, ..., {\bf r}_N \})= \sum_{i>j} v(|{\bf r}_i - {\bf r}_j|),
\end{equation}
where the dependence of the potential energy on the cluster configuration has been emphasized.
The configuration of a cluster which minimizes the potential energy function can be denoted by a set of coordinates
$\{{\bf r}^0_1, {\bf r}^0_2, ..., {\bf r}^0_N \}$ which denote mean, static positions of the $N$ particles 
within the cluster. Assuming 
that ${\bf u}_i = {\bf r}_i - {\bf r}^0_i; |{\bf u}_i| \ll |{\bf r}^0_i - {\bf r}^0_j|, \forall i,j \in \{1,...,N \}$, i.e. that the displacements of particles from their equilibrium positions are small, one can expand
the total potential energy of the cluster in the Taylor series up to the second order as, 
\begin{equation}
V_p (\{{\bf r}_1, ..., {\bf r}_N \}) = V_p(\{{\bf r}^0_1, ..., {\bf r}^0_N \}) + \frac{1}{2}\sum_{i,j} \sum_{\alpha , \beta} {u}_i^{\alpha} {u}_j^{\beta} \left( \frac{\partial ^2 V_p}{\partial {r}_i^{\alpha} \partial {r}_j^{\beta}} \right) _0,
\label{eq:qharm}
\end{equation}
where $\alpha$ and $\beta$ denote the Cartesian components ($x,y$, and $z$) of the vectors.
The first derivatives of $V_p$ with respect to atom coordinates are assumed to vanish, i.e. the 
cluster is assumed to be in a minimum potential energy configuration. This is the well known harmonic approximation and it serves as a starting point for the 
calculation of the cluster normal modes of vibration \cite{Born}, i.e. a set of linear combinations of 
$\{ {\bf u}_1, ... {\bf u}_N \}$ variables (or eigenmodes), each of which corresponds to a vibration 
of the system with a single frequency \cite{Goldstein}. 
The Hamiltonian of the problem when written in terms of 
normal mode coordinates represents a set of independent harmonic oscillators whose both 
quantum and classical dynamics are well known. There are $3N-6$ such 
oscillators with characteristic frequencies $\omega_p , p=1,...,3N-6$. Six degrees of freedom 
that do not represent vibrations are three rotations and three translations of the whole system.

Once a set of eigenfrequencies is calculated, one can proceed to calculate the Helmoltz free 
energy of the cluster, $F$, which is given by
\begin{equation}
F = -k_B T \ln Z,
\end{equation}
where $k_B$ is the Boltzmann constant, $T$ is temperature, and $Z$ is the quantum partition function 
of a system of $3N-6$ independent oscillators. In terms of the eigenmode frequencies, one 
can write \cite{Born}
\begin{equation}
F = V_{p}^0 + \sum_{p=1} ^{3N-6} \frac{\hbar \omega_p}{2} + k_B T \sum_{p=1} ^{3N-6} \ln \left [ 1 - \exp \left ( \frac{\hbar \omega_p}{k_B T} \right ) \right ],
\label{eq:free2}
\end{equation}
where
\begin{equation}
V_p^0 = V_p(\{{\bf r}^0_1, ..., {\bf r}^0_N \}),
\end{equation}
is the minimum of classical potential energy of the cluster (classical ground state energy), 
and $\hbar$ is the reduced Planck 
constant. The sum of the first two terms in Eq. (\ref{eq:free2}) represents the quantum ground 
state energy of the cluster, $E_0$, calculated in the harmonic approximation.

At constant temperature, the state which represents thermodynamical equilibrium of the system 
is the one which minimizes the Helmholtz free energy \cite{Huang}. Note that even at zero 
temperature, the 
cluster free energy has a quantum, zero-point energy contribution [second term in Eq.(\ref{eq:free2})], 
in agreement with the ideas put forth in the Introduction. Thus, even at zero temperature, 
the thermodynamical equilibrium state of the cluster need not be the same as the state 
which minimizes the classical potential energy of the cluster \cite{Doye2}. This of course depends 
on the 
nature of particles in the system, their mass in particular as is known from the studies of 
systems of He atoms \cite{He_clust}.

\section{Potential energy minima, vibrational spectra and zero-point energies of 
Lennard-Jones CSIC and CSFCC clusters}
\label{sec:sec2}

In order to perform a normal vibrational mode calculation, one first has to find the set of 
coordinates $\{{\bf r}^0_1, {\bf r}^0_2, ..., {\bf r}^0_N \}$ which make the functional $V_p$ 
stationary in the $3N$-dimensional configurational space. If the normal mode calculation were 
performed in a nonstationary configuration, some of the normal mode frequencies 
would turn out imaginary. This would signify the instability of the cluster structure \cite{Born}.

The atoms 
were assumed to interact via binary Lennard-Jones 6-12 potentials,
\begin{equation}
v(|{\bf r}_i - {\bf r}_j|) \equiv v(r) = 4 \epsilon \left[ \left( \frac{\sigma}{r} \right)^{12} 
- \left ( \frac{\sigma}{r} \right)^6 \right],
\end{equation}
and were initially arranged in a configuration which has the symmetry of either cuboctahedral or 
icosahedral cluster with the nearest-neighbor atom distances set close to the Lennard-Jones range 
parameter $\sigma$. The thus obtained initial configuration was then allowed to relax to a configuration 
in which the forces acting on each of the atoms in the cluster were smaller than some predefined and 
arbitrarily small absolute force. To obtain the relaxed configuration, each of the atoms in the 
cluster was displaced a certain distance in the direction of the total force acting on it. This was repeated until the absolute value of the force averaged over all cluster atoms dropped below predefined 
force $f_c$. In each step of this iterative procedure, the lengths of vectors the atoms 
were moved along were reduced or enlarged, depending on the vectorial characters of forces in a given and 
preceding step acting on a particular atom. Apparently similar relaxation algorithms were used in Refs. \cite{Burton,Burton2,McGinty1,Northby1}. The value of $f_c$ used in the calculations presented below was $2.0*10^{-13}$ $\epsilon / \sigma$, except for the $n=2$ CSFCC cluster for which $f_c=2.0*10^{-7}$ $\epsilon / \sigma$ was used. For smaller values of $f_c$, the algorithm used relaxes the cluster of 
$n=2$ CSFCC initial symmetry to the $n=2$ CSIC configuration. Whether this signifies instability of CSFCC $n=2$ cluster, as is the case for 
$n=1$ CSFCC cluster \cite{Burton}, or is simply a manifestation of inability of the algorithm to reach presumably 
shallow and/or narrow minimum in the configurational space, is not clear and is of limited importance to this work - the $n=2$ CSFCC cluster was treated as stable and the calculation of vibrations has been 
performed. 
It suffices to say that the potential energy obtained for $n=2$ CSFCC cluster in this study agrees 
perfectly with the result of the previous study \cite{Xie} (see Table \ref{tab:tab1}). Nevertheless, 
it is also of interest to note that the results of Refs. \cite{Burton2,McGinty1} also indicate the 
instability of $n=2$ CSFCC cluster.

The classical potential energies of the clusters, $V_p^0$ obtained using the procedure explained in 
the previous paragraph are displayed in Table \ref{tab:tab1} for closed-shell clusters with maximum shell 
number $n$. These energies depend only on $\epsilon$, and this was used as a scale for 
$V_p^0$. The minimum nearest neighbor distance within the cluster, 
$r_{min} = \min \{|{\bf r}_i - {\bf r}_j|\}, i,j=1,...,N, i \ne j$, is also displayed.
\begin{table}
\caption{Classical potential energies ($V_p^0$) in units of $\epsilon$ and minimum nearest neighbor distances ($r_{min}$) in units of $\sigma$ 
of Lennard-Jones icosahedral and cuboctahedral closed-shell clusters. The maximum cluster shell number 
is denoted by $n$, and total number of atoms in the cluster by $N$.} 
\begin{ruledtabular}
\begin{tabular}{|c|c|c|c|c|c|}
\multicolumn{2}{c}{-} & \multicolumn{2}{c}{Icosahedral, CSIC} & \multicolumn{2}{c}{Cuboctahedral, CSFCC} \\
\hline
n & N & $V_p^0$ [$\epsilon$] & $r_{min}$ [$\sigma$] & $V_p^0$ [$\epsilon$] & $r_{min}$ [$\sigma$] \\
\hline
2 & 55 & -279.2485 & 1.05045 & -268.2765 & 1.09399 \\
3 & 147 & -876.4612 & 1.03548 & -854.3766 & 1.09093 \\
4 & 309 & -2007.219 & 1.02596 & -1971.561 & 1.08998 \\
5 & 561 & -3842.394 & 1.01904 & -3792.097 & 1.08929 \\
6 & 923 & -6552.723 & 1.01361 & -6488.217 & 1.08894 \\
7 & 1415 & -10308.89 & 1.00914 & -10232.14 & 1.08872 \\
8 & 2057 & -15281.55 & 1.00535 & -15196.07 & 1.08842 \\
9 & 2869 & -21641.35 & 1.00205 & -21552.22 & 1.08821 \\
\end{tabular}
\end{ruledtabular}
\label{tab:tab1} 
\end{table}
The values of $V_p^0$ are equal to those obtained in Ref. \cite{Xie} in the first six significant 
decimal places. This confirms the validity of the relaxation algorithm used in this work. The 
minimal nearest neighbor distance found in clusters is always between the central atom and 
one of the atoms in the first cuboctahedral or icosahedral shell. It is of interest to note 
that the nearest neighbor distance in Lennard-Jones 6-12 FCC crystal is $1.09017 \sigma$ \cite{Born}.

When the equilibrium configuration was obtained, the force-constants 
[second derivatives of the binary potential function, see Eq.(\ref{eq:qharm})]
acting between all pairs of atoms within the cluster were calculated, the dynamical vibrational matrix was 
set up and its diagonalization performed in order to obtain eigenmode frequencies and polarization 
vectors \cite{Born}. The frequencies obtained in such a way depend on the atom mass, $M$, and on the Lennard-Jones 
potential parameters, but only through their combination, $\omega_0 \equiv \sqrt{\epsilon / (M \sigma ^2)}$ 
\cite{deWette}. This combination of parameters was used as a universal frequency scale.

Figures \ref{fig:fig2} and \ref{fig:fig3} display the vibrational densities of states per atom, $\rho (\omega)/N$ of CSIC and CSFCC clusters, respectively, as a function of maximum cluster shell number. 
The vibrational density of states was calculated as
\begin{equation}
\rho (\omega) = \sum_{p=1}^{3N-6} \delta(\omega-\omega_p).
\end{equation}
For the sake of easier visualization, the $\delta$-functions in the above equation were represented by 
normalized gaussians with a width parameter of 0.02 $\omega_0$. The 
phonon density of states per atom of FCC Lennard-Jones 6-12 crystal is displayed in the 
bottom panels of Figs. \ref{fig:fig2} and \ref{fig:fig3} for comparison. The phonon density 
of states for FCC crystal was calculated by numerical sampling the of three-dimensional 
inverse (wave vector) space in 3000000 randomly distributed points.

One can immediately note quite different scales on $x$-axes in Fig. \ref{fig:fig2} 
and Fig. \ref{fig:fig3}. This is due to the fact that the vibrational spectra of CSIC clusters contain 
a high-frequency tail which is not present in the CSFCC case. For example, the maximum frequency of 
$n=3$ CSIC cluster is $\omega_{max} = 37.20 \omega_0$, while in 
$n=3$ CSFCC cluster it amounts to $\omega_{max} = 25.96 \omega_0$. 
The reason for such a large difference can be found in the data presented in Table \ref{tab:tab1}.
From the inspection of minimum nearest-neighbor distances, one can conclude that the CSIC 
clusters are more tightly packed and that some of the neighbors are in a repulsive 
region of their binary interaction potential (the minimum of the Lennard-Jones 6-12 
binary potential is at $r=1.122 \sigma$ \cite{sibphon}). This especially applies to the central cluster 
atom which is very tightly surrounded by the atoms in the first icosahedral shell (see Table 
\ref{tab:tab1}). The modes that dominantly represent relative 
motion of such atoms have therefore quite high frequencies. To substantiate this claim, 
I have plotted in Fig. \ref{fig:fig4} the eigenvectors (or displacement patterns) of some of the characteristic 
modes pertaining to CSIC clusters. The displacement pattern of the highest 
frequency mode of $n=2$ CSIC cluster is represented in panel (b) of Fig. \ref{fig:fig4}. It 
can be seen that in this mode the motion of the central atom dominates the displacement pattern. 
Atoms in the first shell also slightly move, so that the total linear and angular momenta of the 
mode equal to zero, as they should. Analogous mode in the $n=3$ CSFCC cluster is depicted in panel (b) of 
Fig. \ref{fig:fig5}. Again, the highest frequency mode is such that the central atom 
performs motion with the largest amplitude of all atoms in the cluster. The displacement 
patterns of the lowest frequency 
modes in $n=2$ CSIC and $n=3$ CSFCC clusters are depicted in panels (a) of Fig. \ref{fig:fig4} and 
\ref{fig:fig5}, respectively. It is of interest to note that the displacement pattern of 
the lowest frequency mode in $n=2$ CSIC cluster is a sort of a "twisting" mode in which two halves 
of the cluster perform motions which look almost like the rotations around the same axis, but in 
opposite directions for the "upper" and "lower" halves of the cluster.

Further inspection of the vibrational densities of states reveals significant differences between 
CSFCC and CSIC clusters even in the region of frequencies which contains the highest percentage 
of all vibrational modes (i.e. disregarding the high-frequency tail of the density of states in 
CSIC case). While $\rho(\omega)$ of CSFCC clusters obviously tends to the bulk 
(crystal) limit (except for the characteristic features around 10.8 $\omega_0$), the same does not 
hold for $\rho(\omega)$ of CSIC clusters which behaves quite differently even for the largest 
cluster considered ($n=9$). This was illustrated by a superposition of $\rho(\omega)$ for CSIC and 
CSFCC $n=9$ clusters in panel denoted by $n=9$ in Fig. \ref{fig:fig3}. Low frequency vibrations 
(up to about 8 $\omega_0$) are quite similar in both types of sufficiently large clusters, but 
CSIC density of states does not exhibit characteristic features around 25 $\omega_0$, which 
in a FCC crystal (bottom panel) are a consequence of the van Hove type of singularity related to 
features of dispersion of 
longitudinal and transversal modes (phonons) at the crystal Brillouin zone edges (see e.g. Ref. \cite{Gupta}).

The origin of the features around 10 $\omega_0$ in CSIC and 10.6 $\omega_0$ CSFCC 
clusters is in the large number of surface (poorly coordinated) atoms \cite{Meyer}. Even for the largest 
cluster considered, made of 2869 atoms, 812 atoms are located at the cluster surface which makes about 
28 percent (for smaller clusters the percentage is higher). This means that about 28 percent 
of the vibrational density of states represents the modes in which surface atoms are significantly 
displaced from their equilibrium positions. Such modes are not present in the bulk crystal calculation. 
The peaks in $\rho(\omega)$ can in fact be related to the zone-edge frequencies of 
the Rayleigh-wave (RW) modes of surfaces of Lennard-Jones 
crystals \cite{deWette,sibxe}.
Additional confirmation of the identification of characteristic peaks can be found by inspecting 
panel (c) of Fig. \ref{fig:fig4} which displays the displacement pattern of a mode of 
$n=4$ CSIC cluster with frequency of 10.33 $\omega_0$. In this particular mode, mostly the surface atoms 
vibrate and the polarization vectors are oriented dominantly perpendicularly to the cluster surface. A  similar pattern was found for larger CSIC and CSFCC clusters. However, the 
quality of the visual insight in the polarization pattern tends to degrade with the number of 
atoms in the cluster and this was the main reason for the choice of relatively small clusters 
for a visualization of the displacement patterns in Figs. \ref{fig:fig4} and \ref{fig:fig5} (see 
panel (c) in Fig. \ref{fig:fig4}).

In Table \ref{tab:tab2}, the zero point energy, 
\begin{equation}
F_0 = \frac{1}{2}\sum_{p=1} ^{3N-6} \hbar \omega_p 
\end{equation}
pertaining to the CSFCC and CSIC clusters is presented.
\begin{table}
\caption{Zero point energies ($F_0$) in units of $\hbar \sqrt{\epsilon / (M \sigma ^2)}$ of Lennard-Jones 
icosahedral and cuboctahedral closed-shell clusters. The maximum cluster shell number 
is denoted by $n$.} 
\begin{ruledtabular}
\begin{tabular}{|c|c|c|}
- & Icosahedral, CSIC & Cuboctahedral, CSFCC \\
\hline
n & $F_0$ [$\hbar \sqrt{\epsilon / (M \sigma ^2)}$] & $F_0$ [$\hbar \sqrt{\epsilon / (M \sigma ^2)}$] \\
\hline
2 & 1018.571 & 1049.012 \\
3 & 3083.370 & 3173.793 \\
4 & 6854.732 & 7007.818 \\
5 & 12787.93 & 13210.08 \\
6 & 21138.61 & 22359.53 \\
7 & 33097.20 & 34923.92 \\
8 & 48960.54 & 51457.37 \\
9 & 68577.49 & 72560.07 \\
\end{tabular}
\end{ruledtabular}
\label{tab:tab2} 
\end{table}
These results are to some extent surprising. Although both the 
minimum and maximum frequency are smaller in the CSFCC clusters, the first moments of their 
frequency spectra are {\em larger} than in corresponding CSIC clusters, at least for the 
cluster shell numbers between 
2 and 9 studied here. One can visually inspect that this is indeed so by looking again at 
the panel denoted by $n=9$ in Fig. \ref{fig:fig3}. This means that quantum ground state energies 
differ more than 
(classical) potential energies of the clusters. Thus, icosahedral closed shell clusters are 
even more preferred energetically when the quantum nature of the particles is important (in the cases 
when one can speak about the ordered ground state, when the approach presented here is adequate).
To illustrate this effect and estimate its magnitude for clusters composed of rare gas atoms, 
in Fig. \ref{fig:fig6} I plotted the difference between the classical potential energies of CSIC and 
CSFCC clusters [$\Delta V_p^0 = V_p^0$(CSIC)$- V_p^0$(CSFCC), full circles] and the difference 
between the quantum 
ground state energies [$\Delta E_0 = V_p^0$(CSIC)$- V_p^0$(CSFCC) + $F_0$(CSIC)$- F_0$(CSFCC), empty diamonds] as a function of cluster shell number $n$, and for clusters composed of Ne, Ar, and Xe atoms. 
These quantities cannot be obtained for general Lennard-Jones clusters in some reduced units 
of energy since the classical potential energy scales with $\epsilon$, while the zero-point 
energy scales with $\hbar \omega_0$.
The Lennard-Jonnes parameters used in this calculation were $\epsilon (Ne)=3.07$ meV, 
$\sigma (Ne)=2.75$ \AA, $\epsilon (Ar)=10.35$ meV, $\sigma (Ar)=3.40$ \AA, 
$\epsilon (Xe)=19.18$ meV, $\sigma(Xe)=4.1$ \AA \cite{liquid}. This produces characteristic 
frequency scales $\hbar \omega_0(Ne)=0.29$ meV, $\hbar \omega_0(Ar)=0.31$ meV, and 
$\hbar \omega_0(Xe)=0.19$ meV. The classical results (full circles) are practicaly the 
same as those obtained in Fig. 2 of Ref. \cite{Xie}. It is obvious that quantum 
corrections become more important for lighter atoms. Even for Xe clusters, their effect is 
not negligible, especially when the differences between the CSIC and CSFCC clusters are 
considered. Intriguingly enough, the relative importance of quantum effects increases with 
the shell number. The behavior of free energy with temperature is such that the differences 
between the CSIC and CSFCC free energies increase in absolute value as the temperature 
increases. This means that at finite temperatures, the CSIC structure is even more 
favored than at $T=0$ K. The behavior of the sum of second two terms in Eq. (\ref{eq:free2}), 
\begin{equation}
F_1 \equiv F - V_p^0 = \sum_{p=1} ^{3N-6} \frac{\hbar \omega_p}{2} + k_B T \sum_{p=1} ^{3N-6} \ln \left [ 1 - \exp \left ( \frac{\hbar \omega_p}{k_B T} \right ) \right ],
\label{eq:f1}
\end{equation}
with temperature is illustrated in Fig. \ref{fig:fig7} for $n=5$ CSFCC (full line) and CSIC 
(dotted line) clusters. Similar behavior is found for all $n$'s examined in this work. 

\section{Summary and Conclusion}
\label{sec:sec3}

The main result of this article is that quantal treatment of the low temperature properties 
of 2 to 9 shell CSIC and CSFCC Lennard-Jones clusters results in a larger binding energy of CSIC 
clusters. The difference between binding energies of CSIC and CSFCC clusters is larger than 
in classical treatment \cite{Xie}. This results is somewhat surprising in the view of expectations 
put forth in the Introduction. It is plausible, although not shown in this article, that the presented quantal treatment would yield a quantum binding energy "crossover" between the CSFCC and CSIC clusters for 
larger shell numbers than obtained classically (14), i.e. that the CSFCC 
arrangement of clusters would become energetically favorable for larger clusters than predicted 
previously \cite{Xie}. On the basis of Fig. \ref{fig:fig6} it can be expected that the exact 
number of atoms for which the crossover takes place is dependent on the mass of atoms in the 
cluster.

\begin{figure}[h]
\caption{Closed shell icosahedral (CSIC, top panel) and cuboctahedral (CSFCC, bottom panel) 
Lennard-Jones clusters 
in their minimum potential energy configuration for maximum shell numbers $n=$2,3, and 4 (from 
left to right).}
\label{fig:fig1}
\end{figure}

\begin{figure}[h]
\caption{Vibrational densities of states of CSIC clusters as a function of maximum shell 
number, $n=2,...,9$. The vibrational density of states of Lennard-Jones FCC crystal is displayed 
in the bottom panel for comparison.}
\label{fig:fig2}
\end{figure}

\begin{figure}
\caption{Vibrational densities of states of CSFCC clusters as a function of maximum shell 
number, $n=2,...,9$. The vibrational density of states of Lennard-Jones FCC crystal is displayed 
in the bottom panel for comparison. In panel denoted by $n=9$, the vibrational density of 
$n=9$ CSIC cluster is also displayed and denoted by a dotted line.}
\label{fig:fig3}
\end{figure}

\begin{figure}
\caption{Eigenvectors (displacement patterns) of three vibrational modes pertaining 
to CSIC clusters. The modes depicted in panels (a), (b), and (c) are those denoted by a, b, and 
c in Fig. \ref{fig:fig2}, respectively. Equilibrium positions of cluster atoms are denoted 
by small cubes. The displacement 
vectors are multiplied by 10 in panels (a) and (b), and by 20 in panel (c).}
\label{fig:fig4}
\end{figure}

\begin{figure}
\caption{Eigenvectors (displacement patterns) of some of the vibrational modes pertaining 
to CSFCC clusters. The modes depicted in panels (a), (b), and (c) are those denoted by a, b, and 
c in Fig. \ref{fig:fig3}, respectively. Equilibrium positions of cluster atoms are denoted 
by small cubes. The displacement vectors are multiplied by 15 in panels (a),(b), (c).}
\label{fig:fig5}
\end{figure}

\begin{figure}
\caption{The difference between the classical (full circles) and quantum (empty diamonds) 
ground state energies of CSIC and CSFCC clusters as a function of maximal shell number for Ne (top panel), Ar (middle panel), and Xe (bottom panel) clusters.}
\label{fig:fig6}
\end{figure}

\begin{figure}
\caption{Free energies of $n=5$ CSFCC (full line) and CSIC (dotted line) clusters measured from 
the corresponding classical potential energy minima [see Eq. (\ref{eq:f1})] as a function 
of reduced temperature $k_B T$.}
\label{fig:fig7}
\end{figure}

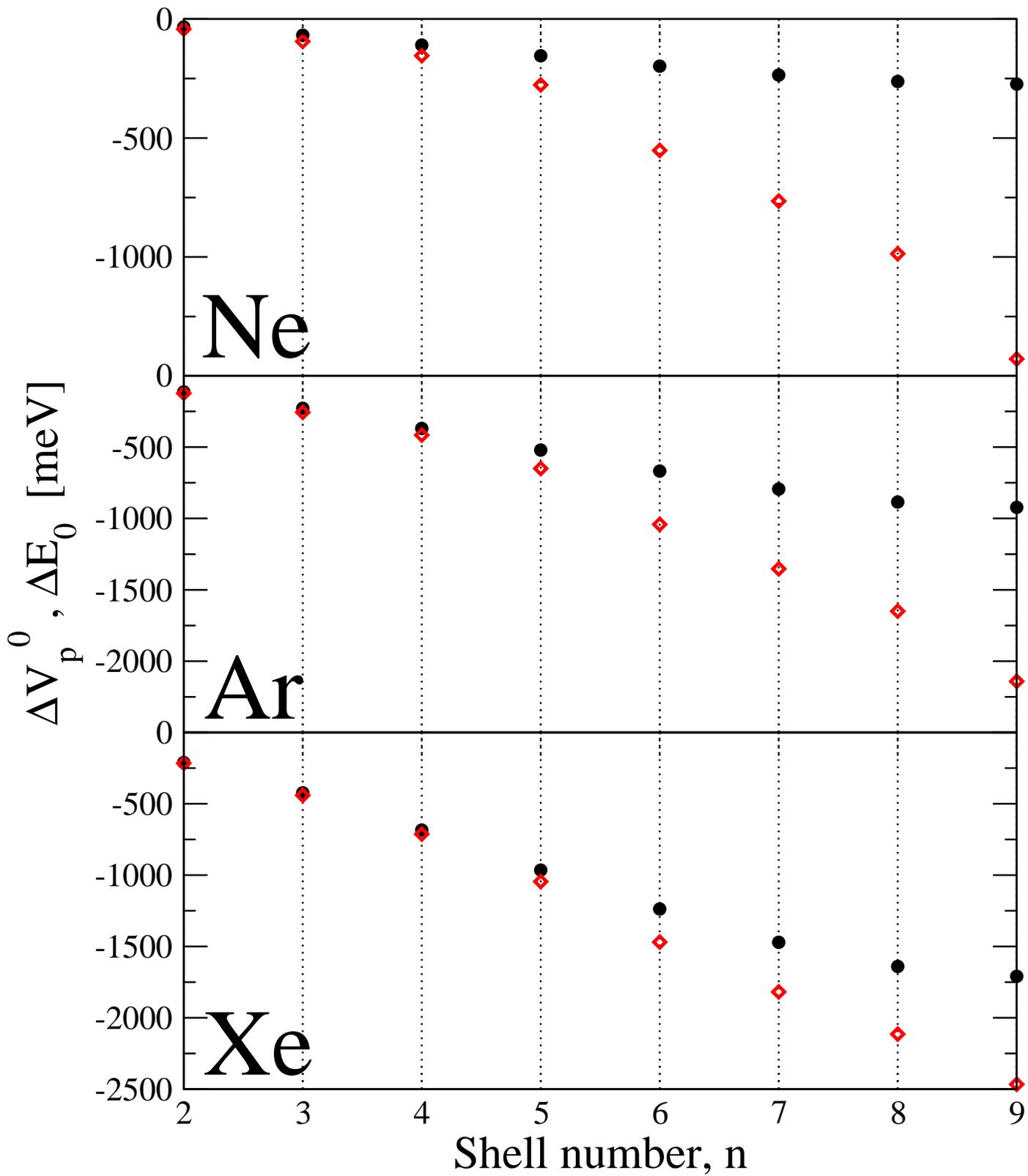
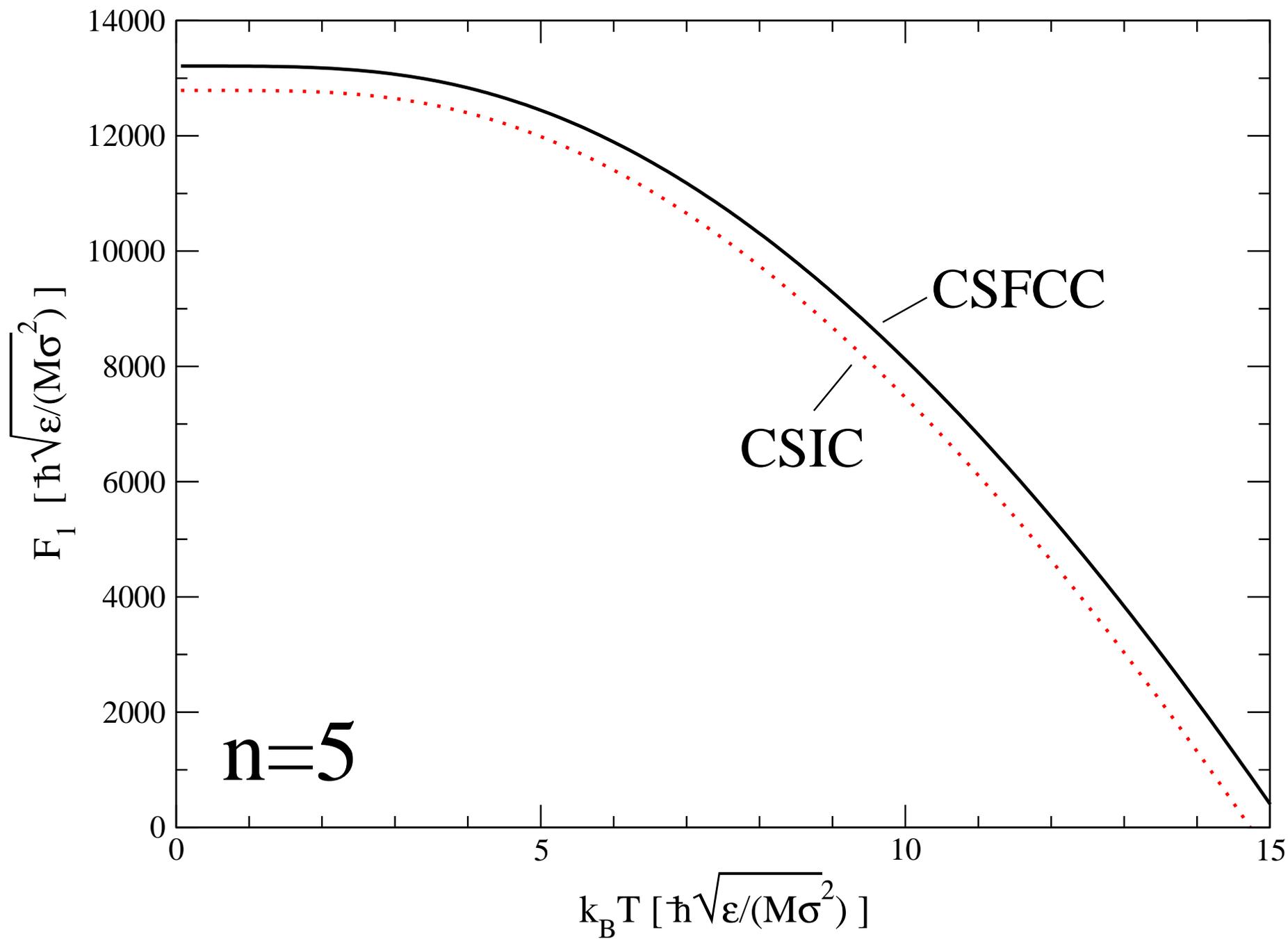
\end{document}